\documentclass[12pt,preprint]{aastex}

\usepackage{lscape}

\def\stacksymbols #1#2#3#4{\def\theguybelow{#2}
        \def\verticalposition{\lower#3pt}
        \def\spacingwithinsymbol{\baselineskip0pt\lineskip#4pt}
        \mathrel{\mathpalette\intermediary#1}}
\def\intermediary #1#2{\verticalposition\vbox{\spacingwithinsymbol
        \everycr={}\tabskip0pt
        \halign{$\mathsurround0pt#1\hfil##\hfil$\crcr#2\crcr
                \theguybelow\crcr}}}
\def\lta{\stacksymbols{<}{\sim}{2.5}{.2}}
\def\gta{\stacksymbols{>}{\sim}{3}{.5}}

\shorttitle{Age of Elliptical Galaxies}
\shortauthors{Temi et al.}

\begin{document}

\title{The Ages of Elliptical Galaxies from Mid-Infrared Emission}

\author{Pasquale Temi\altaffilmark{1,2}, William G. Mathews\altaffilmark{3}, 
Fabrizio Brighenti\altaffilmark{3,4} }

\altaffiltext{1}{Astrophysics Branch, NASA/Ames Research Center, MS 245-6,
Moffett Field, CA 94035.}
\altaffiltext{2}{SETI Institute, Mountain View, CA 94043.}
\altaffiltext{3}{University of California Observatories/Lick Observatory,
Board of Studies in Astronomy and Astrophysics,
University of California, Santa Cruz, CA 95064.}
\altaffiltext{4}{Dipartimento di Astronomia,
Universit\`a di Bologna, via Ranzani 1, Bologna 40127, Italy.}

\begin{abstract}
The mid-infrared (10-20$\mu$m)
luminosity of elliptical galaxies is dominated by
the integrated emission from circumstellar dust in red giant stars.
As a single stellar population evolves, the rate of
dusty mass loss from
red giant stars decreases with time, so the mid-infrared
luminosity should also decline with stellar age.
To seek such a correlation, we have used archival ISO observations
to determine surface brightness profiles and central fluxes
at 15$\mu$m in 17 early-type galaxies for which stellar ages
have been determined from optical spectral indices.
The radial surface brightness distributions at 15$\mu$m
generally follow the stellar de Vaucouleurs profile as expected.
We find that the surface brightness ratio $\mu_{15\mu m}/\mu_{I-band}$
is systematically higher in elliptical galaxies with ages
$\lta 5$ Gyrs and in galaxies that exhibit evidence of
recent mergers.
Within the accuracy of our observations, $\mu_{15\mu m}/\mu_{I-band}$
shows no age dependence for ages $\gta 5$ Gyrs.
The corresponding flux ratios $F_{15\mu m}/F_{I-band}$ within
apertures scaled to the effective radius ($R_e/8$)
are proportional to the $\mu_{15\mu m}/\mu_{I-band}$ ratios
at larger galactic radii,
indicating that no 15$\mu$m emission is detected from
central dust clouds visible in optical images
in some of our sample galaxies.
Emission at 15$\mu$m is observed in non-central massive clouds
of dust and cold gas in
NGC 1316, an elliptical galaxy that is thought to have had
a recent merger. Recent {\it Spitzer Space Telescope} data also indicate
the presence of PAH emission at $8\mu$m.
Several ellipticals have extended regions
of 15$\mu$m emission
that have no obvious counterparts at other frequencies.

\end{abstract}


\section{Introduction}

Recently \citet{pio03} studied the
mid- and far-infrared emission from an evolving
stellar population formed as a burst with a Salpeter IMF.
Infrared emission at $\sim9-20\mu$m arises from
warm dust in the stellar winds of evolved giant stars;
emission at shorter wavelengths $\lta 9\mu$m can
be largely photospheric
\citep{ath02}.
Most of the time variation of mid-infrared emission from an
evolving stellar population results from the
declining rate of stellar mass loss from red giants,
but \citet{pio03} include many additional details.
The mid-infrared emission is far less sensitive to
stellar metallicity than age.
This raises the interesting possibility that the mean
stellar ages in elliptical galaxies
can be determined directly from mid-infrared observations.
We explore this possibility here using archival ISO observations
at 15$\mu$m.

The mean ages of stars in elliptical galaxies can be
estimated optically from the strength of the
stellar Balmer lines
(e.g. Gonzalez 1993, Worthey 1994,
Tantalo et al. 1998, Trager et al. 2000, Kuntschner et al. 2002).
These mean age determinations, accurate to 20-30 percent,
are often interpreted by comparing optical spectra
with computed time-dependent
spectral energy distributions a single burst of star formation.
Most of the stars in elliptical galaxies are thought to be
very old, possibly $\sim 10$ Gyrs,
but unexpectedly young
ages have been found for many ellipticals (e.g. $\lta 5$ Gyrs).
Evidently the mean stellar age has been skewed downward by the
addition of a relatively small component of young stars
that may result from a recent merger or internal
star formation.

Using archival ISO data at 15$\mu$m for 17 early-type galaxies,
mostly ellipticals, we explore the possibility that
warm circumstellar dust can be used to determine stellar
age or the likelihood of a recent merger.
Balmer line stellar ages are known for all galaxies in our sample.
We begin by verifying that the surface brightness profiles at
15$\mu$m are similar to those in the I-band which we assume to
be stellar, unaffected by dust emission.
Since optically obscuring dust is often observed in the cores
of elliptical galaxies, including some in our sample,
we used both the integrated (within $R_e/8$) flux ratio
$F_{15\mu m}/F_{I-band}$ and
the surface brightness ratio $\mu_{15\mu m}/\mu_{I-band}$
at a larger radius ($R_e/4$) as a measure of the
mean stellar ages. The two methods give similar results.
In fact we find that the flux ratio $F_{15\mu m}/F_{I-band}$ within
$R_e/8$ is similar to the
ratios $\mu_{15\mu m}/\mu_{I-band}$ at $R_e/4$ for all the galaxies
in our sample.
Apparently, optically obscuring dust in the galactic cores
-- which could in principle contribute to the emission at 15$\mu$m --
cannot be easily detected at the sensitivity
and resolution of ISO.
As we show below, a plot of $F_{15\mu m}/F_{I-band}$ against
optically determined ages shows that the mid-infrared emission is
indeed sensitive to stellar age provided the galaxies are
no older than $\sim 5$ Gyrs.
Abnormally large $F_{15\mu m}/F_{I-band}$ or
$\mu_{15\mu m}/\mu_{I-band}$ ratios,
suggesting the presence of younger stars, are found in
elliptical galaxies that (1) are thought to have suffered a recent
merger (2) have strong optically obscuring dust,
or (3) have HI or non-thermal radio emission.

\section {Observations}

The infrared camera ISOCAM on board the ISO satellite (Kessler et
al. 1996)
observed about 60 elliptical galaxies in the 4-15$\mu$m region with
narrow and
wide band filters.
>From these observations we selected 22 early-type galaxies,
17 of which have
optically determined stellar
ages (Terlevich \& Forbes 2002).
The relevant data for the galaxies in our sample are listed 
in Table 1.
For these sample galaxies we consider only those observations
taken with the smallest pixel scale to maximize the accuracy of
the radial and azimuthally averaged mid-infrared emission.
Mid-infrared fluxes within small radii are required to detect
emission from dust in the galactic cores.
The selected observations were performed in raster mode in
the LW3 ($\lambda_{ref} = 14.3\mu$m) filter.
Individual galactic images have been taken with the $32 \times 32$
Si:Ga array camera with 3$^{\prime \prime}$ pixel scale.
The total on-source integration times for the Si:Ga array
are typically 300-600 seconds.
Each observation consists of a collection of exposures,
each taken with the target galaxy centered
at several positions on the array.
Two galaxies, NGC 4552 and IC 1409, have been observed at a
plate scale of $6^{\prime \prime}/$pix and with larger pointing
offsets
among the individual exposures, allowing raster maps
of approximately $3.5 \times 9^{\prime}$.

\section {Data Reduction}

Data reduction has been performed using the Camera Interactive
Analysis (CIA) package, version 5.0 \citep{ott97}.
The procedure to reduce ISOCAM data starts with the dark current
subtraction, for which we used the {\it model} option to take
into account the long term drift in the dark current which
occurred during the ISO mission. Then we removed cosmic
ray hits using a sigma clipping filter or multi-resolution
method. This removes most glitches of short duration.
We also visually inspected the frames to manually
remove glitches with very long time constants that may affect
more than one frame.

Si:Ga devices are also affected by transient effects
in which the response time to any change in illumination
depends on the incident photon flux.
This transient behavior needs to be characterized
for our specific data set
and properly allowed for in recovering
the ``stabilized'' electronic signal.
The transient correction was performed using
the Fouks-Schubert method \citep{cou00} in which a model of the
detector response to incident radiation is used
to reconstruct the stabilized fluxes.
We then flat-fielded the data
using standard library flats and applied the conversion
from electronic units to Jy/pix
using the standard {\it cal--g} calibration data
that accompanies the data products.
After each single exposure
has been properly reduced, the final image was produced by registering
and coadding each frame.

All 22 galaxies were detected at 15$\mu$m with a S/N in excess
of $\sim10$. In all cases the field of view was larger
than the size of the galaxy, allowing an evaluation
of the background flux directly from the galaxy maps.
The sky background was calculated by taking the median
pixel value from several blank regions in the reduced frames.
We used the reduction routines in the IRAF package
to perform both the aperture photometry
and the azimuthally averaged radial profiles.

\section {Results}

One of the goals in this study is to test the
sensitivity of mid-infrared
surface brightness profiles to the age of the old stellar
population in elliptical galaxies
using the known age measurements from optical studies.
In principle it might be possible to determine the
stellar age at any galactic radius
by comparing the surface brightness profiles
at 15$\mu$m and optical wavelengths.
Our measurements of the 15$\mu$m surface brightness profiles
may allow us to detect emission from
dust whose spatial distribution does not follow
the galactic starlight.
Complementary
observations of the 15$\mu$m flux through a small aperture
centered on the galactic core are designed to detect
emission from optically obscuring dusty clouds
observed near the centers of many elliptical galaxies
in our sample.

\subsection{Radial Surface Brightness Profiles at 15$\mu$m}

We used the task {\it ellipse} in the IRAF {\it stsdas} package
to perform elliptical isophote analysis.
The isophote fitting algorithm, described by
\citet{jed87}, computes best-fitting elliptical isophotes.
It applies iterative corrections
to the geometrical parameters of a trial ellipse
by projecting the fitted harmonic amplitude
onto the image plane.
The output provides all geometrical parameters for the ellipse
including the semi-major axis $R$ and mean isophotal
intensity.
The left panels in Figure 1
show the 15$\mu$m surface brightness profiles
for each galaxy as a function of the semi-major axis to the
1/4 power.
Most galaxies have surface brightness profiles
$\mu_{15\mu{\rm m}}(R)$ that are generally in very good
agreement with a de Vaucouleurs profile,
which is a straight line in these plots.
The downward curvature of $\mu_{15\mu{\rm m}}(R)$ at small $R$
arises because the pixel size becomes too large to properly track
the steep de Vaucouleurs profiles there.
It has been recognized for some time that
mid-infrared emission from relatively warm dust
in elliptical galaxies closely follows the
de Vaucouleurs $R^{1/4}$ profile
of optical starlight
\citep{kna92, ath02, xil04},
indicating that the warm dust is photospheric or
circumstellar.
The arrows in the profile plots show the location of
$R_e/8$ for each galaxy where $R_e$ is the effective radius.

The right panels in Figure 1 show the ratio of the
surface brightness at 15$\mu$m to that in the I--band
where dust emission is not expected.
I--band profiles (not shown here) are taken from
the studies of \citet{idi02} and \citet{gou94b}.
To avoid complications when comparing the 15$\mu$m and I-band profiles
due to sampling with very different instruments and pixel scales,
we discarded
the innermost part of the profile of
$\mu_{15\mu m}/\mu_{I-band}$ in Figure 1.
The radial profiles for $\mu_{15\mu m}/\mu_{I-band}$
are also truncated
at large radii where the low S/N precludes a proper isophotal fit.
In the range plotted -- where high S/N, low geometrical distortion,
and reasonably large number of pixels ensure the most reliable
data -- the ratio $\mu_{15\mu m}/\mu_{I-band}$
is approximatively constant with radius for most of the galaxies.
Although the stellar metallicity (and dust production?)
is expected to decrease with
galactic radius, this trend is not reflected in the
$\mu_{15\mu m}/\mu_{I-band}$ profiles, which are either flat
or increase slowly with projected radius.
This insensitivity to stellar metallicity is consistent
with the models of \citet{pio03}.
We note that \citet{mal00}
observe decreasing 15$\mu$m/7$\mu$m with galactic radius,
but dust may contribute at both these wavelengths.

Two galaxies -- NGC 5866 and NGC 5044 -- have
discordant 15$\mu$m surface brightness profiles.
NGC 5866 is an approximately edge-on S0 galaxy with an extensive
dusty disk.
The increase in $\mu_{15\mu m}/\mu_{I-band}$
for $0.9 > R^{1/4} > 1.1$
visible in Figure 1 is due to a slope change in $\mu_{I-band}$
that presumably arises from the bulge-disk transition.
Azimuthally averaged surface brightness profiles for this
galaxy are of limited value.
It is debatable if this dusty S0 galaxy should be included in our
sample of elliptical galaxies, but we have retained it
to see if the dust is visible at 15$\mu$m.
The $\mu_{15\mu m}$ profile for NGC 5044 also has a feature visible
in Figure 1 at $R^{1/4} = 1.15$ caused by asymmetric
emission toward the east and southeast that is discussed
in more detail below.

\subsection{Aperture Photometry}

At visual wavelengths stellar ages and metallicities
in elliptical galaxies
are evaluated from an analysis of spectral indices,
assuming single stellar populations (SSP).
Since line index strengths in early-type galaxies
have radial gradients \citep{car93, dav93, gon93},
the representative mean ages and metallicities
of elliptical galaxies refer to a fixed aperture.
Following \citet{ter02}
and other authors, we compare 15$\mu$m and I-band
fluxes within an aperture of radius $R_e/8$.

Photometric data reduction of the 15$\mu$m maps has been performed
using the {\it apphot} package
in IRAF. These routines are well suited
to compute accurate centers and integrated fluxes within the
specified circular aperture.
Some galaxies in our sample have relatively small
effective radii, requiring an accurate centering algorithm.
It is also essential to account for
pixels that span across the $R_e/8$
aperture to retrieve reliable photometry just within $R_e$/8.

Whenever possible,
I-band fluxes within $R_e/8$ for our sample galaxies
are found by integrating published
$\mu_{15\mu m}(R)$ profiles to this radius.
For galaxies without published $\mu_{15\mu m}(R)$ profiles
we used the total apparent corrected I-magnitude
from the Lyon/Meudon Extragalactic Database (LEDA).
Then, assuming a de Vaucouleurs I-band distribution,
we integrated over the $R^{1/4}$ profile to
determine the flux within the $R_e/8$ aperture using
the appropriate effective radius $R_e$ for each galaxy.
The aperture and surface photometry results are presented
in Table 2.
Figure 2 shows the $F_{15\mu m}/F_{I-band}$ ratios based on the
photometry inside $R_e$/8 plotted against the average stellar
population age of the galaxies as tabulated by
\citet{ter02} for this same aperture. Most optical ages have errors
less than 20\%; age errors are not plotted for clarity. 
The lines in Figure 2 show
the expected evolution of $F_{15\mu m}/F_{I-band}$
produced by SSP models from \citet{pio03}
for metallicities Z = 0.02 (solar), Z = 0.04  and Z = 0.008.
The predicted flux ratios $F_{15\mu m}/F_{I-band}$
from \citet{pio03} are very insensitive to metallicity variation
and vary only weakly for stellar ages $\gta 5$ Gyr.
The observed $F_{15\mu m}/F_{I-band}$ for galaxies with
ages $\gta 5$ Gyr suggest that the variation with stellar
age is even less than that predicted by \citet{pio03} and
that the normalization of the theoretical curves
may be a bit too low at ages $\sim 10$ Gyrs.
Recently \cite{tho04} have provided
independent optical
ages for a large number of the galaxies in our sample, using
models for stellar absorption indices computed with variable
stellar $\alpha$/Fe abundance ratios.
While their ages tend to be slightly larger than those quoted
by \citet{ter02}, our conclusions are unaffected 
if galactic ages from \cite{tho04} were used instead. 
Some of the galaxies with ages $\gta 5$ Gyr are known to contain
central dust clouds and disks (see Table 1),
but there is only a limited correspondence
with the incidence of optically visible dust and the position of
galaxies in Figure 2.

To illustrate the lack of significant 15$\mu$m emission from
central dust, we plot in Figure 3 the relationship between
$R_e/8$ aperture flux ratios
$F_{15\mu m}/F_{I-band}$
and the corresponding surface brightness ratios
$\mu_{15\mu m}/\mu_{I-band}$ evaluated at $R_e/4$,
regarded as the ratio for the stellar emission alone.
If central dust clouds contribute to the 15$\mu$m emission, this
would be revealed in $F_{15\mu m}/F_{I-band}$ but
not in $\mu_{15\mu m}/\mu_{I-band}$.
The data points in Figure 3 lie along the 1:1 line with very
little scatter, suggesting that the vertical spread of the data
showed in Figure 2 cannot be explained by central dust clouds.
One of the elliptical galaxies with the largest positive
deviation in Figure 3 is NGC 5831 which contains no dust visible
with HST.
If small, centrally located dust is present in these galaxies,
its effect on the 15$\mu$m emission is not prominent.

For stellar ages $\lta 3-5$ Gyr
the $F_{15\mu m}/F_{I-band}$ curves in Figure 2 from \citet{pio03}
are very sensitive to stellar age
and in general terms this is supported by our observations.
Three galaxies in Figure 2 -- NGC 5866, NGC 6776 and NGC 1316 --
lie far above the predicted lines for their optically determined ages.
Three additional galaxies -- NGC 1453, NGC 4278 and NGC 4261 --
are marginally higher.
We have already mentioned the unusual S0 galaxy NGC 5866, a galaxy
with a massive dusty disk of cold gas that has been detected in HI and
CO emission (Roberts et al. 1991; Welch \& Sage 2003),
so it is not surprising that it has an additional
component of non-circumstellar mid-infrared dust emission.
Of the remaining galaxies, four -- NGC 1316, NGC 1453, NGC 4261
and NGC 4278 -- show evidence of recent mergers,
have significant HI (which is unusual for
elliptical galaxies) or have strong radio emission
and dusty disks.
However, the remaining galaxy, NGC 6776, seems to be rather normal
but does have a faint shell that may be merger-related.
On balance, therefore, it appears that 15$\mu$m observations can
be used to identify post-merger ellipticals but the
theoretical SSP flux ratios plotted in Figure 2 may
unfortunately be too simplistic
to fully describe the complexities of post-merger dust formation.
The agreement of NGC 5018 with the \citet{pio03} predictions
may be fortuitous since there is some evidence of HI and radio
emission from this galaxy.
However, it appears to be significant that
all four of the ellipticals with the lowest values of
$F_{15\mu m}/F_{I-band}$ in Figure 2 -- NGC 5831, NGC 720,
NGC 2300 and NGC 4649 -- have been observed with HST and found
to contain no visible central dust.

\section{Non-elliptical 15$\mu m$ Images}

In Figure 4 we show 15$\mu$m images of six galaxies from our sample
that have low luminosity extensions in various directions from the
emission peak which is coincident with the optical galaxy.
With a  pixel scale of 3$^{\prime \prime}$ the ISOCAM beam size
in the LW3 filter is approximately 5$^{\prime \prime}$ FWHM.
Thus, some of the features in the central 30$^{\prime \prime}$
of NGC1316 and NGC1453 are clearly resolved.
In the remaining galaxies, other non-axisymmetric
emission features are apparent
but are not resolved by these observations.
Apart from NGC 5044, the azimuthally averaged 15$\mu$m surface
brightness profiles tend to wash out the faint isophotal
extensions visible in Figure 4.
In general, we have been unable to find emission at other
frequencies that correlates with these mid-infrared features.
The exception to this is NGC 1316 (Fornax A) a galaxy thought to
have had recent merger (Schweizer 1980;
Geldzahler \& Fomalont 1984).
Dust and gas appear to be falling into NGC 1316, forming a
disk-like pattern $\sim15^{\prime \prime}$ in radius in the
SE-NW direction, roughly parallel to the jet radio axis
\citep{gel84}.
The extended emission in our 15$\mu$m image is entirely consistent
with emission from optically visible
dust in these same regions, confirming the
interpretation of \citet{xil04}.

Recently the Spitzer Space Telescope \citep{wer04} has observed NGC1316
with the IRAC \citep{faz04} instrument in four mid-infrared bands at 3.6, 4.5,
5.8 and
8.0$\mu$m. The higher sensitivity and spatial resolution
of the camera allow us to better resolve the emission in
the central $2^\prime$ of the galaxy.
The 8$\mu$m image in Figure 5 (left panel) shows in great detail 
and at higher S/N the same asymmetrical features present
in the ISO 15$\mu$m
image (see also \cite{xil04}).
Many bright knots and an arc-like
structure are prominent and well resolved by IRAC.

The right panel in Fig. 5 shows a contour plot of the emission
at 8.0$\mu$m once the stellar continuum is subtracted
using a scaled 3.6 $\mu$m image. The non--stellar emission is then
superimposed to a B--I HST image (grayscale) (press release
STScI-PR99-06, Grillmair et al. 1999) where the dust distribution
is shown in dark extended features. The correspondence
of the non--stellar 8$\mu$m
emission features with the dust seen in absorption by HST is
remarkable. Almost every feature visible in the optical absorption
study, included the central one, has 8$\mu$m emission associated.

The 8.0$\mu$m broadband filter sample the emission band
centered at ``7.7$\mu$m'' from  Polycyclic Aromatic
Hydrocarbons (PAH) molecules. Much of the excess emission
in this band must then be attributed to PAHs, but the coincidence
of the bright spots with the 15$\mu$m image in the region
where the Spitzer and ISO data overlap is indicative that
warm small dust grains also contribute to the mid-infrared emission.

Perhaps young, luminous stars are forming in the infalling dusty gas,
raising the dust temperature sufficiently to emit at 15$\mu$m.
It is interesting that dust located at large radii 
in NGC1316 is associated
with 15$\mu$m emission, while optically visible dust in the
galactic cores is not.
This may occur because only stars of rather low
(Jeans) mass can form
near the galactic cores where the gas pressures are very high.
But the mass of molecular hydrogen in the dusty patches
along the central SE-NW direction in NGC 1316,
$5 \times 10^8$ $M_{\odot}$
\citep{hor01},
exceeds by $\sim 100$ the mass of dusty cold gas
typically observed in elliptical galaxy cores,
and this may be the best explanation for the associated
star formation.
Finally, it seems odd that the low luminosity extensions in
the 15$\mu$m image of NGC 5044, which we believe to be real,
are strong enough to alter the surface brightness profile but
still have no easily identifiable correlated emission at other
frequencies.

\section{Conclusions}

We have found that the infrared emission at 15$\mu$m from elliptical
galaxies, normalized by the I-band emission,
can provide an approximate measure of the mean age of galactic stars.
Similar to the predictions of \citet{pio03}, the
central flux ratios $F_{15\mu m}/F_{I-band}$
generally increase for
(1) ellipticals known to have stellar ages $\lta 5$ Gyrs and
particularly for (2) ellipticals
that exhibit unusual attributes associated with recent mergers:
dusty disks, HI emission, non-thermal radio emission, shells, etc.
However, if only normal, non-merging galaxies are considered,
the central flux ratios $F_{15\mu m}/F_{I-band}$
for galaxies in our sample shows little or no
convincing variation with optically determined stellar ages.
Since reliable estimates of the dust mass in elliptical galaxies
are not available, it is not possible to quantify the excess
15$\mu$m emission expected from a given amount of dust.
However, the four ellipticals having the lowest observed values of
$F_{15\mu m}/F_{I-band}$ all appear to be dust-free in HST
observations.

Some individual galaxies seem inconsistent with these general trends.
NGC 6776 has an unusually large $F_{15\mu m}/F_{I-band}$, but
only modest optical evidence for a recent merger.
NGC 7626 and NGC 4374 have optically obscuring central dust and
double radio lobes but rather normal $F_{15\mu m}/F_{I-band}$.

The strong correlation between the flux ratio $F_{15\mu m}/F_{I-band}$
measured within $R_e/8$ and $\mu_{15\mu m}/\mu_{I-band}$ evaluated
at $R_e/4$ indicates that 15$\mu$m
emission from optically visible dust in the cores of
some of our sample galaxies is not detectable with ISO.
This contrasts with the more distant dust in the merging galaxy
NGC 1316 which is a local source of 15$\mu$m emission.
Evidently, massive dust-heating stars are forming in NGC 1316,
but not in the dusty cores of many other ellipticals.
As in NGC 1316,
five additional elliptical galaxies show non-stellar
extended features in their 15$\mu$m images.
While we believe these features are real, they do not appear associated
with emission at other wavelengths.

\vskip.4in
We thank L. Piovan for providing electronic form of the
SSP model outputs.
Studies of hot gas and dust in elliptical galaxies
at UC Santa Cruz are supported by
NASA grants NAG 5-8409 \& ATP02-0122-0079 and NSF grants
AST-9802994 \& AST-0098351 for which we are very grateful.
PT is supported in part by NASA/ADP grant 21-399-20-01. 
FB is supported in part
by a grant MIUR/PRIN 0180903.
This research is based on observations with ISO,
an ESA project with instruments funded by ESA Member States
and with the participation of ISAS and NASA.
Part of this work is based  on observations made with the
{\it Spitzer Space Telescope}, which is oparated by
Jet Propulsion Laboratory, California Institute of 
Technology, under NASA contract 1407.

\clearpage
\begin{figure}
\plotone{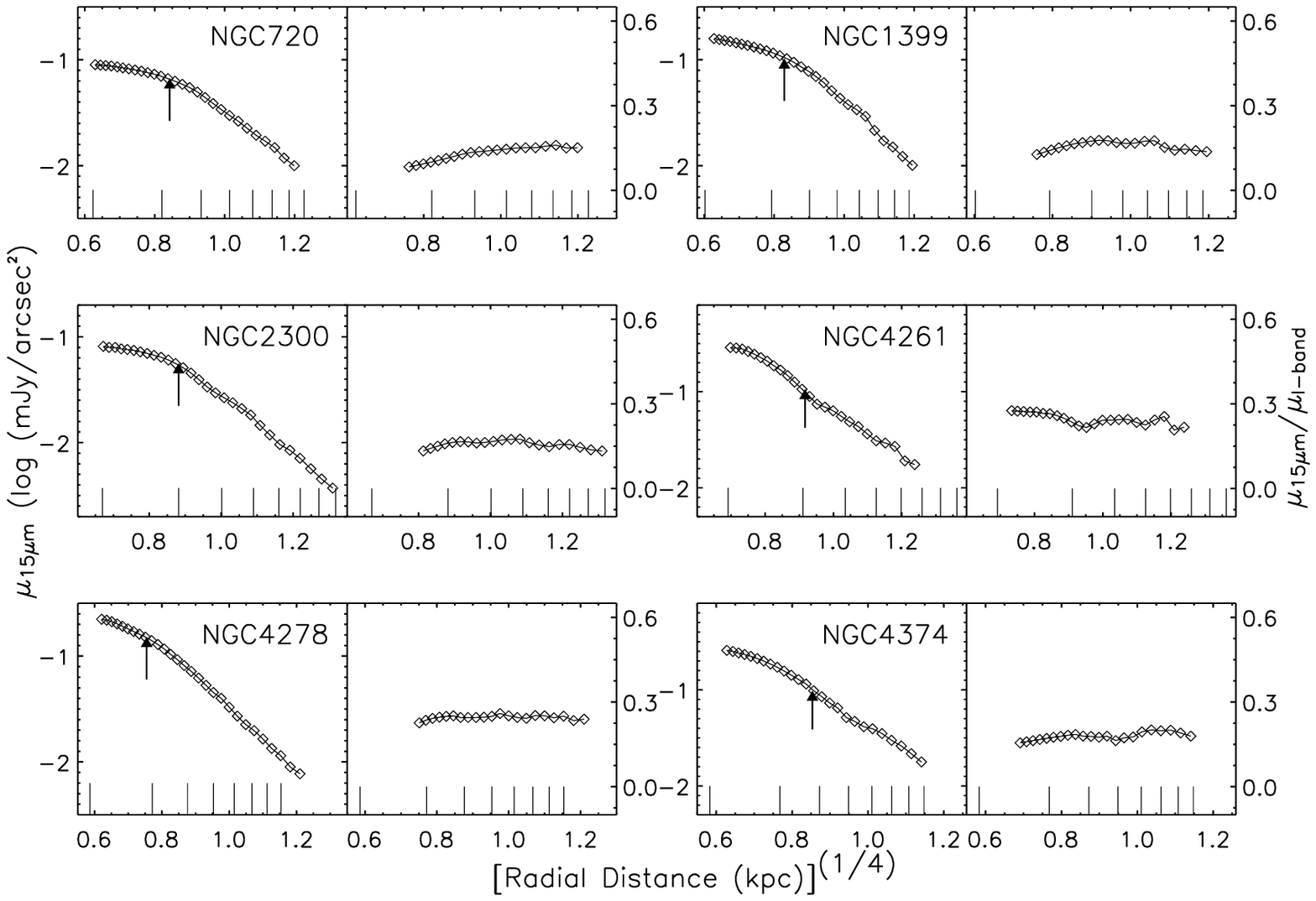}
\caption{{\it Left panels:} Surface brightness profiles at 
15$\mu$m for each galaxy. The arrows 
indicate $R_e/8$ and the vertical lines on the abscissa mark 
off individual pixels; the first bar correspond to the 1/2 pixel position.
{\it Right panels:} the 
radial variation of the surface brightness 
ratio $\mu_{15\mu m}/\mu_{I-band}$. 
}
\end{figure}
\clearpage
\begin{figure}
\plotone{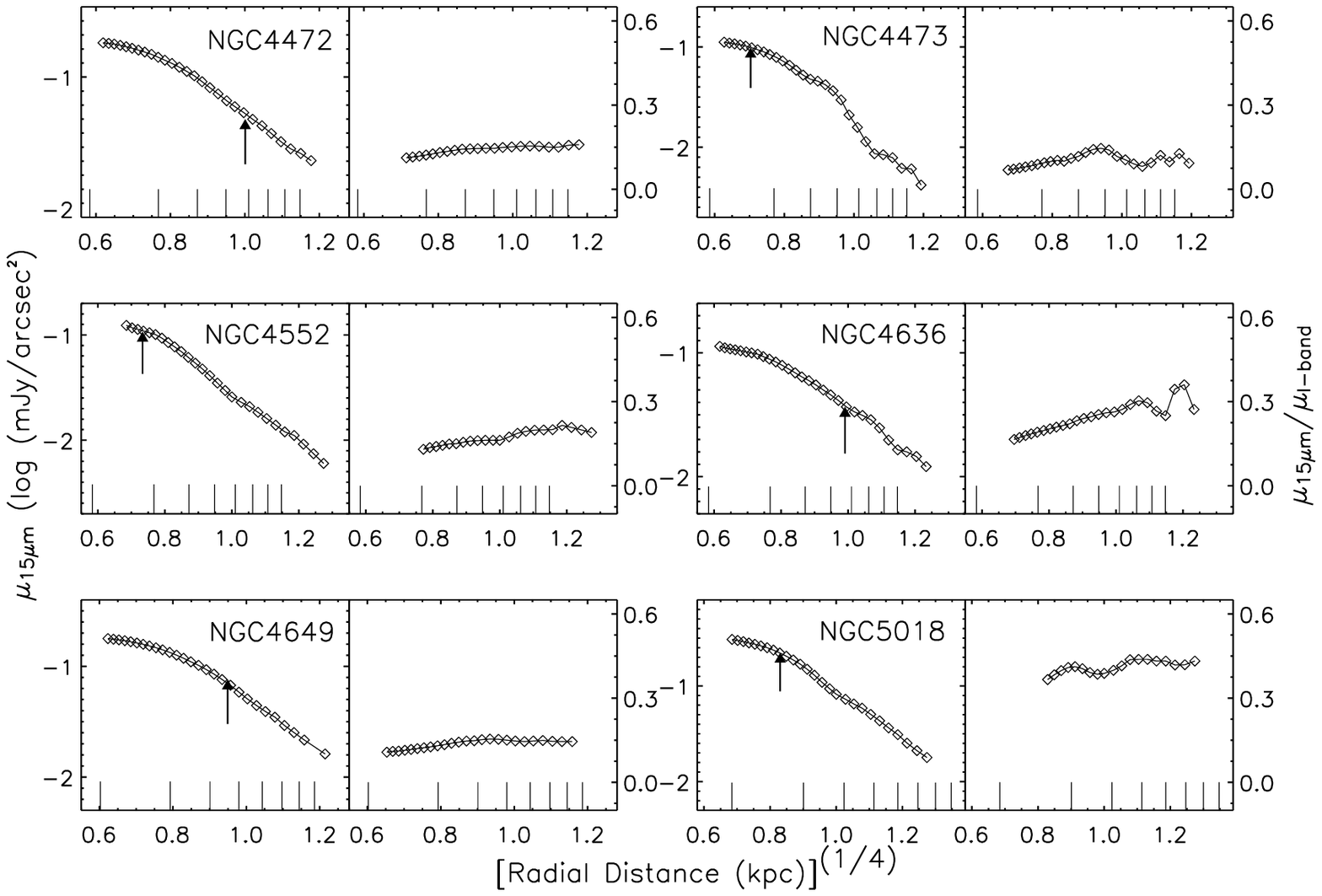}
\end{figure}
\clearpage
\begin{figure}
\plotone{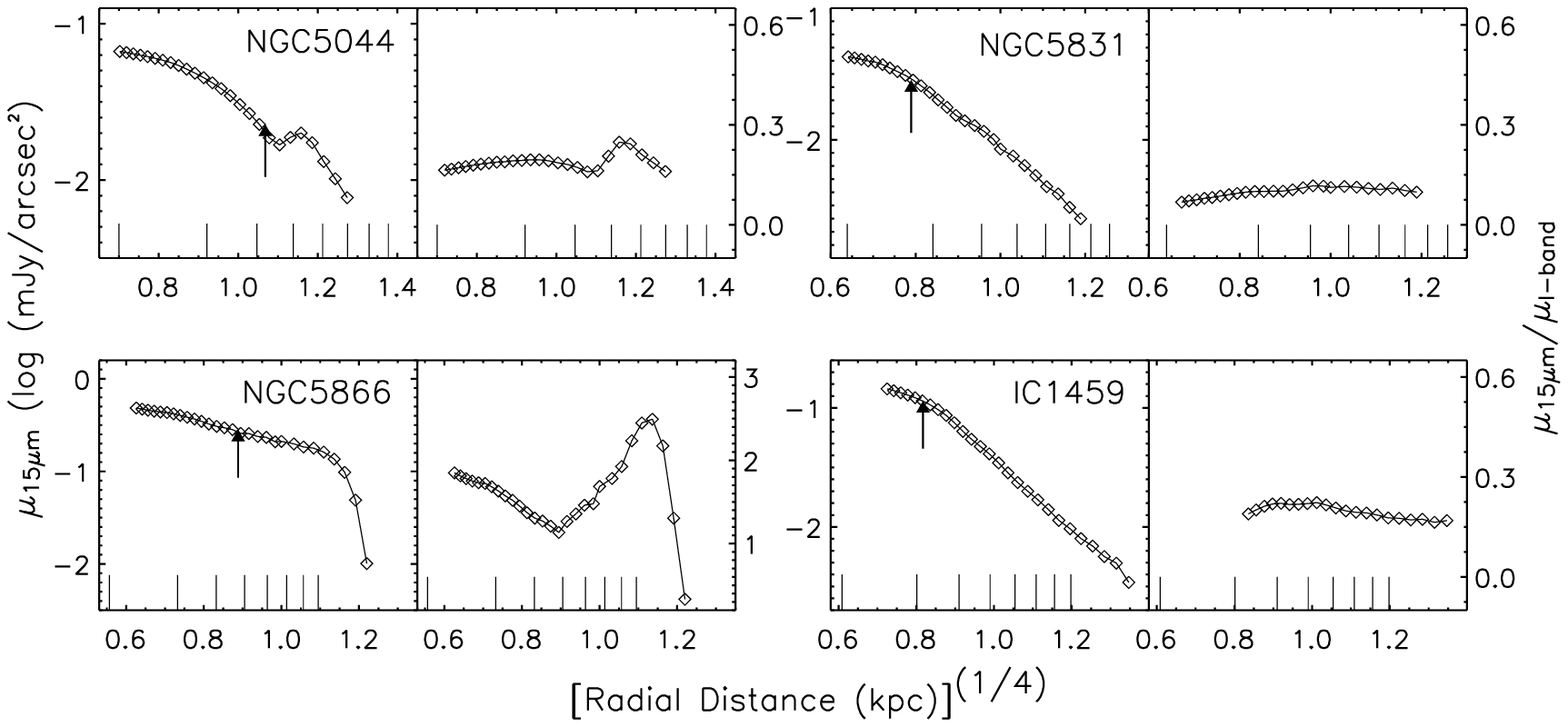}
\end{figure}

\begin{figure}
\plotone{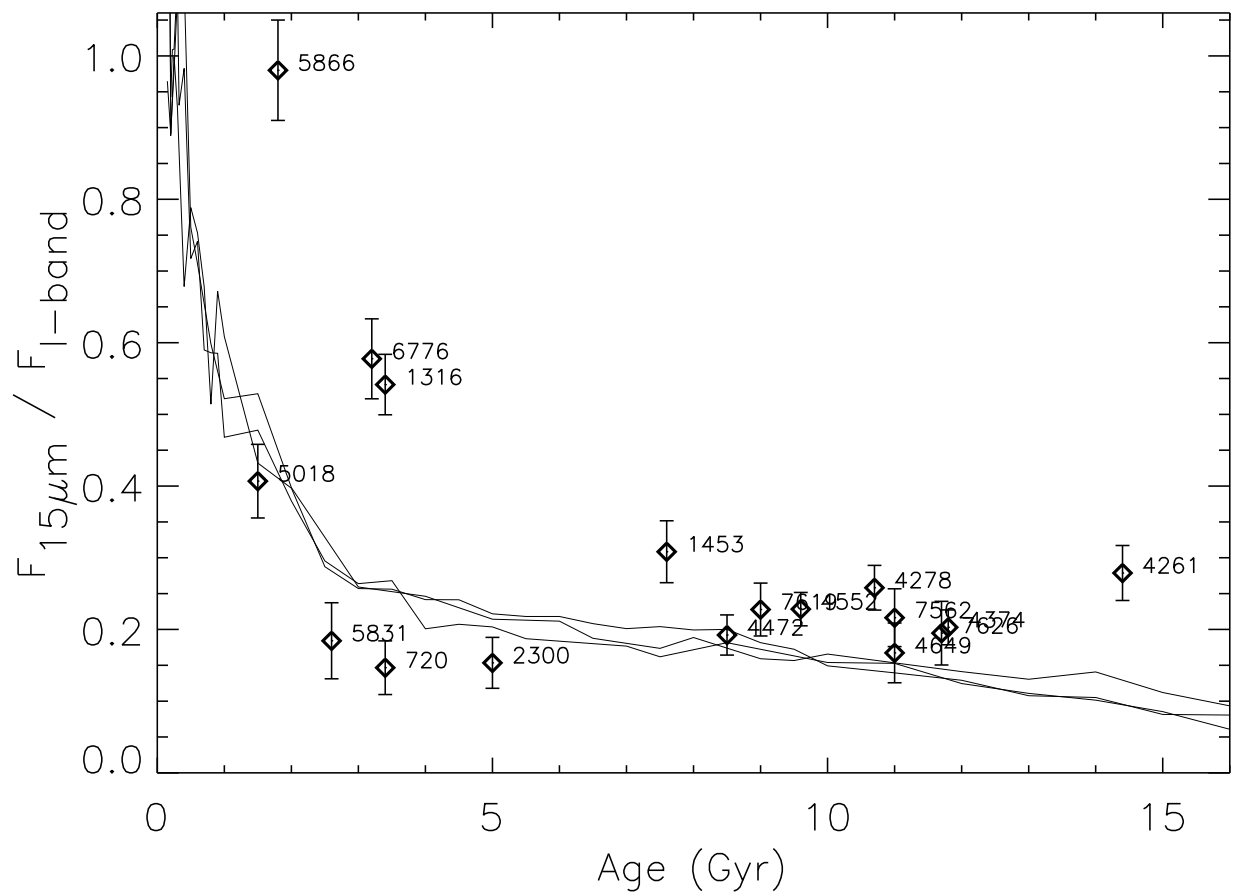}
\caption{The ratio $F_{15\mu m}/F_{I-band}$ within $R_e/8$ 
plotted against optically determined ages 
for each galaxy. The lines show the  
$\mu_{15\mu m}/\mu_{I-band}$ predicted by the single stellar 
population models of \citet{pio03}. 
These overlapping lines for populations with abundances of $Z = 0.008$, 
0.004, and $0.02 = Z_{\odot}$ show that $\mu_{15\mu m}/\mu_{I-band}$
is essentially independent of the stellar abundance. 
Typical errors in the optically determinated  mean ages are $\sim$ 20-30 percent.
}
\end{figure}
\clearpage
\begin{figure}
\plotone{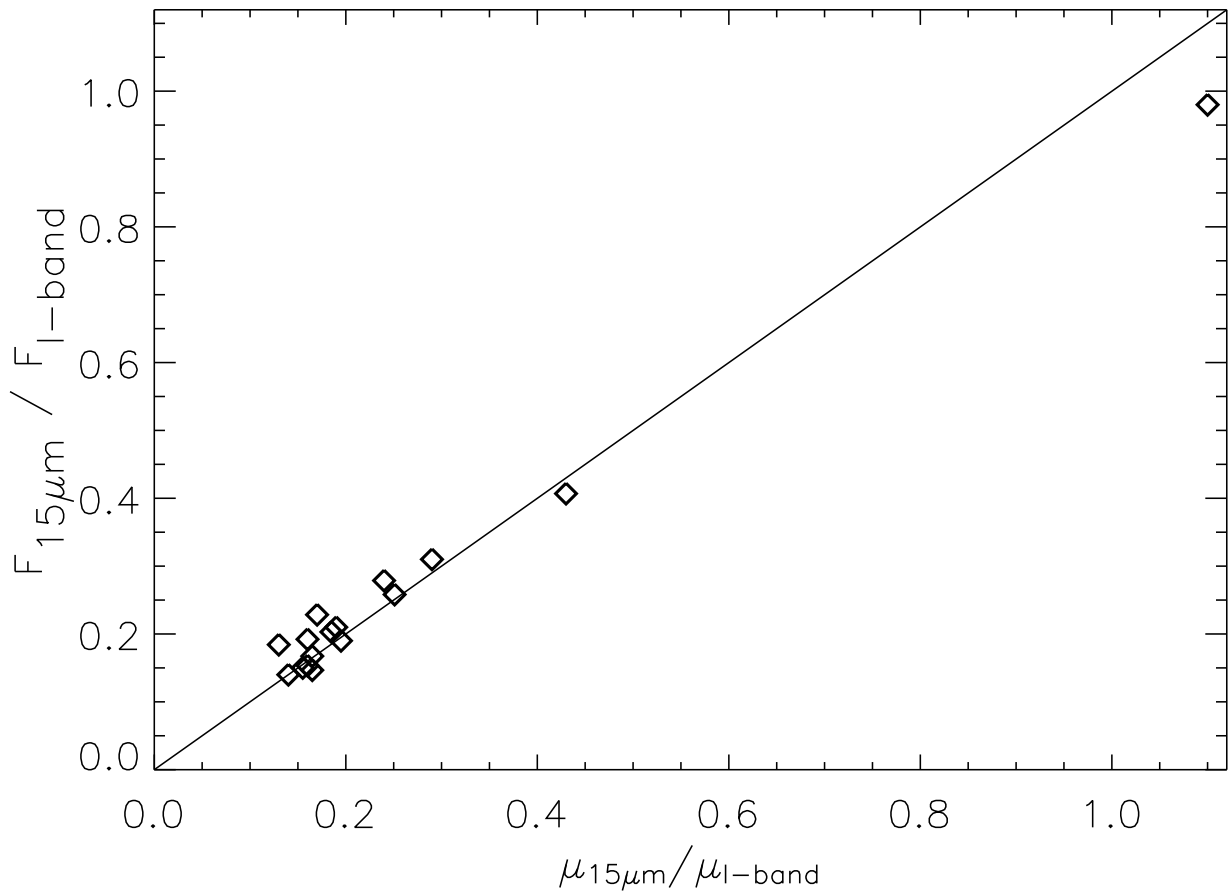}
\caption{Flux ratios $F_{15\mu m}/F_{I-band}$ measured within
apertures of size $R_e/8$ plotted against surface brightness ratios
$\mu_{15\mu m}/\mu_{I-band}$ evaluated at $R_e/4$. We plot values
for the 16 galaxies with measured $I$-band profiles taken from
Goudfrooij et al. (1994) and Idiart et al. (2002). 
Except for one galaxy (NGC5866), a very good linear correlation
is seen between the two presented ratios. The values are consistent,
within the errors, with the 1:1 line.}
\end{figure}

\clearpage
\begin{figure}
\plotone{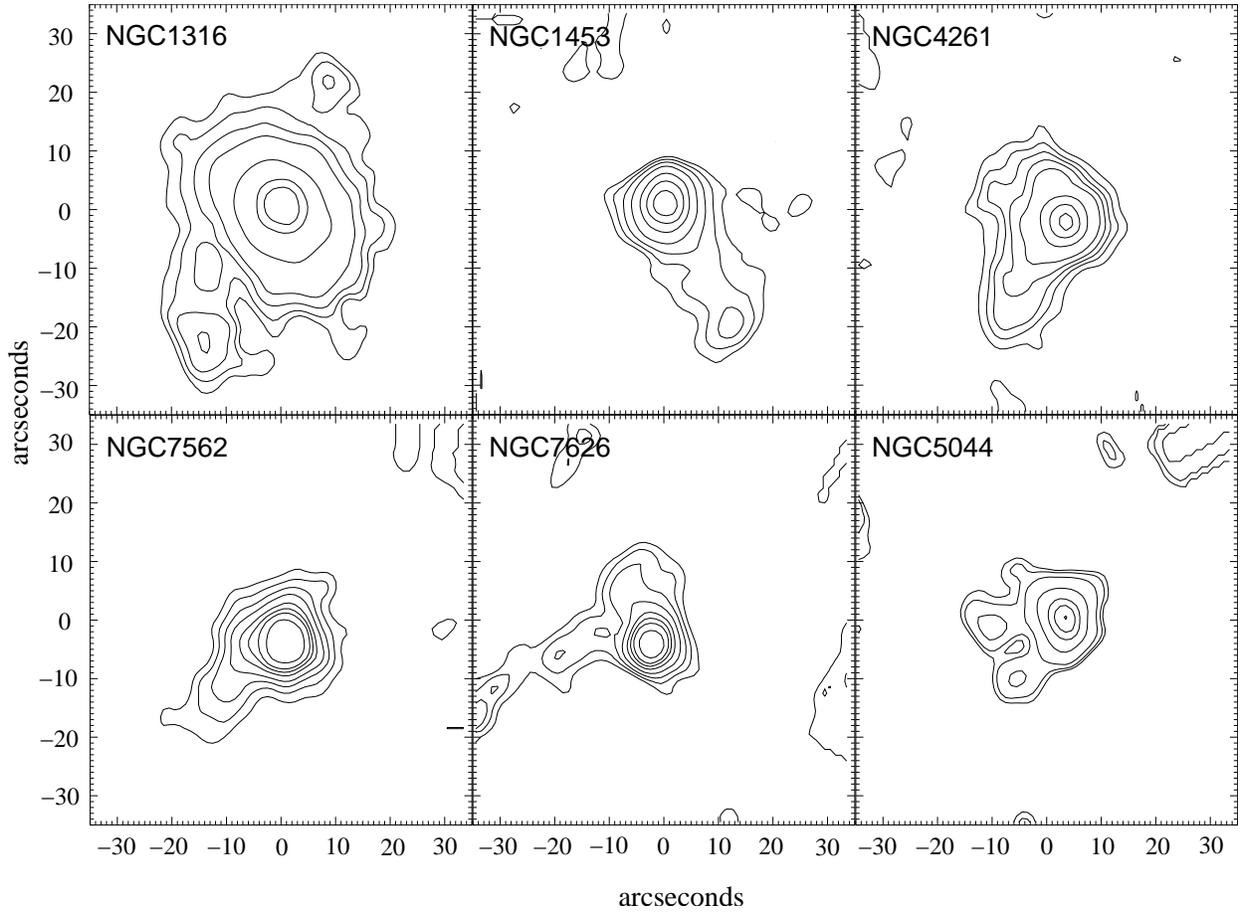}
\caption{15$\mu$m isophotal contours of the central 
$30^{\prime \prime}$ of the six galaxies in our sample 
that show faint, resolved non-symmetric mid-infrared 
extensions in various directions. 
The contour levels in mJy/pixel are: 
0.62, 0.69, 0.79, 0.94, 1.5, 3.0, 4.3 (NGC1316);
0.32, 0.35, 0.39, 0.45, 0.55, 0.66, 0.78, 0.9 (NGC1453);
0.28, 0.37, 0.44, 0.52, 0.65, 1.2, 1.9, 2.5 (NGC4261);
0.2, 0.23, 0.26, 0.29, 0.33, 0.36, 0.39, 0.45 (NGC7562);
0.17, 0.20, 0.23, 0.27, 0.32, 0.36, 0.40, 0.47 (NGC7626);
0.09, 0.11, 0.15, 0.27, 0.40, 0.48, 0.56, 0.65 (NGC5044);
}

\end{figure}

\clearpage
\begin{figure}
\plotone{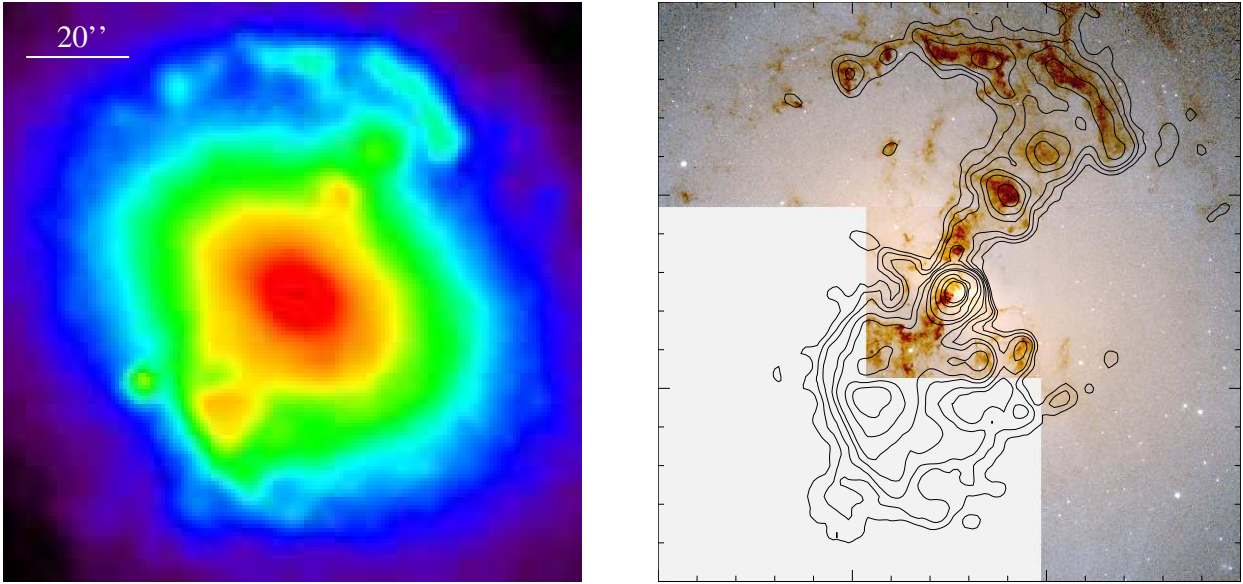}
\caption{NGC 1316 at visual and mid-infrared light. The 8$\mu$m image 
taken with the Spitzer Space Telescope (left panel) shows several 
bright knots, and an arc-like structure at larger distance from
the center. 
The right panel shows an isophotal contour plot of the emission 
at 8$\mu$m with the stellar continuum subtracted, superimposed on 
an HST  B--I image (grayscale) of the central region of NGC 1316.
(press release STScI-PR99-06, Grillmair et al. 1999).
The field of view is $1.9^{\prime} \times 1.9^{\prime}$
with North pointing up and East to the left for both images. 
The dark dust absorption features seen in visual light
are a perfect match for the emission features at 8$\mu$m attributed 
to PAH molecules.  }
\end{figure}

\clearpage


\begin{deluxetable}{lcccccccll}
\setlength{\tabcolsep}{0.02in}
\tabletypesize{\tiny}
\tablewidth{0pt}
\tablecaption{Basic Properties of the Sample}
\tablecolumns{10}
\tablehead{
\colhead{Name}                          &
\colhead{D\tablenotemark{a}}            &
\colhead{ T\tablenotemark{b}}        &
\colhead{ Type\tablenotemark{c}}           &
\colhead{$R_e$}                         &
\colhead{Log $L_{B}$\tablenotemark{a}}  &
\colhead{age \tablenotemark{d}}         &
\colhead{Optical dust}                  &
\colhead{Visible dust}                  &
\colhead{Notes}     \\
\colhead{}                    &
\colhead{(Mpc)}               &
\colhead{}                    &
\colhead{}                    &
\colhead{(Kpc)}   &
\colhead{($L_{B\odot}$)}      &
\colhead{(Gyr)}               &
\colhead{from extintion?}            &
\colhead{lane or disk?}                  &
\colhead{}   \\
}

\startdata

NGC0720........    & 20.80   &   -4.8  & E5          & 4.03& 10.38  &  3.4  &\nodata&no\tablenotemark{9} & a prototype E5 galaxy central galaxy in Fornax ;  \\
                   &         &         &             &     &        &       &       &  & cluster\tablenotemark{27} \ \ ; strong color gradient in stars; central  \\ 
                   &         &         &             &     &        &       &       &  & color is among blueset of E galaxies\tablenotemark{40} \\ 
NGC1316........    & 18.11   &   -1.7  & SABs(0)     &7.08 & 10.93  &  3.4  &\nodata&yes\tablenotemark{9}& evidence of recent merger in H$\alpha$, dust\tablenotemark{20} \ \ ; $H_2$ detected  \\
                   &         &         &             &     &        &       &       &                    & in dust patches\tablenotemark{21} \ \ ; Fornax A (strong radio  source)\tablenotemark{26} \ \ ; \\
                   &         &         &             &     &        &       &       &   & evidence of merger Arp 154\tablenotemark{26, 29} \ \ \ \ \ ; 3 Gyr old merger \\
                   &         &         &             &     &        &       &       &   & remnant\tablenotemark{30} \ \  ; 0.5 Gyr old merger remnant\tablenotemark{33} \ \ .\\ 
NGC1399........    & 18.11   &   -4.5  & E1          & 3.78& 10.52  &\nodata&\nodata&no\tablenotemark{9} & IRAS100$\mu$m\tablenotemark{2} \ ; low luminosity radio jet\tablenotemark{22} \ \ . \\
NGC1453........    & 43.10   &  -5.0   & E2--3       & 5.85& 10.37  &  7.6  &\nodata&\nodata & gas disk\tablenotemark{10} \ \ ; radio\tablenotemark{12} \ \ ; HI\tablenotemark{12, 13} \ \ \ \ \ .   \\
NGC2300........    & 27.67   &  -3.5   & SA0         & 4.83& 10.41  &  5.0  &\nodata&no\tablenotemark{23}& shells; nuclear radio\tablenotemark{2} \ ; a normal E3\tablenotemark{27} \ \ .  \\
NGC4261........    & 31.48   &  -4.8   & E3          & 5.65& 10.70  &  14.4 &\nodata&yes\tablenotemark{2,7,8,9,37}  & powerful double lobe radio (3C270)\tablenotemark{2, 12} \ \ \ \ ; radio jets \\
                   &         &         &             &     &        &       &    &   & perpendicular to disk\tablenotemark{37} \ \ ; strong central optical\\
                   &         &         &             &     &        &       &    &   & emission lines\tablenotemark{38} \ \ ; UV emission from jet\tablenotemark{39} \ \ ; \\
NGC4278........    & 16.22   &  -4.8   & E1          & 2.60& 10.24  &  10.7 &yes\tablenotemark{1}&yes\tablenotemark{2,4,8,36}  & large rotating HI+dust disk\tablenotemark{2,3,12,14,35,37} \hspace{0.6in} ; strong  \\
                   &         &         &             &     &        &       &    &   & nuclear LINER emission lines \tablenotemark{34} \ \ ; core-halo  \\
                   &         &         &             &     &        &       &    &   & radio source with compact core\tablenotemark{35} \ \ ;  \\
                   &         &         &             &     &        &       &    &   & optically visible dust patches\tablenotemark{36} \ \ ; \\
NGC4374........    & 15.92   &  -4.0   & E1          & 4.25& 10.57  &  11.8 &yes\tablenotemark{11}&yes\tablenotemark{2,6,9}   & classical double radio \tablenotemark{2, 12} \ \ \ \ ;  no large scale dust at \\
                   &         &         &             &     &        &       &    &   & submm wavelengths\tablenotemark{24} \ \ ; nuclear dust lane\tablenotemark{37} \ \ ; \\
                   &         &         &             &     &        &       &    &   & core-jet radio source perp. to lane\tablenotemark{37} \ \ ; \\
                   &         &         &             &     &        &       &    &   & one-sided radio "jet"\tablenotemark{35} \ \ ; M84; 3C272.1 \\
NGC4472........    & 15.92   &  -4.7   & E1/S01(1)   &8.03 & 10.90  &   8.5 &yes\tablenotemark{11}&yes\tablenotemark{9}   & radio\tablenotemark{12} \ \ ;  HI\tablenotemark{15} \ \ ; 2-sided weak radio jet\tablenotemark{42} \ \ ; \\
NGC4473........    & 16.14   &  -4.8   & E5          & 1.96& 10.15  &\nodata&\nodata&no\tablenotemark{9} &   very small edge-on dust and gas disk but not see  \\
                   &         &         &             &     &        &       &    &   & with HST\tablenotemark{43} \ \ ; prototypical E5 galaxy\tablenotemark{27} \ \ ;   \\
NGC4552........    & 15.92   &  -4.6   & S01(0)      & 2.32& 10.29  &   9.6 &\nodata&yes\tablenotemark{9}   & $H_{\alpha}$ undetected at $log(L_{H_{\alpha}}) < 38.06$\tablenotemark{41} \ \ ; \\
                   &         &         &             &     &        &       &  &    & modest radio source\tablenotemark{44} \ \ ; M89;  \\
NGC4636........    & 15.92   &  -4.8   & E0/S01(6)   &7.71 & 10.51  & \nodata&\nodata& yes\tablenotemark{9}   &  NICMOS image is smooth, showing no dust  \\
                   &         &         &             &     &        &        & &    &  deviates from fund. plane for E cores\tablenotemark{36} \ \ ; \\
                   &         &         &             &     &        &        & &    &  radio core plus jets\tablenotemark{45} \ \ ;   \\
                   &         &         &             &     &        &        & &    &  moderately strong broad $H_{\alpha}$\tablenotemark{46} \ \ ;  \\
                   &         &         &             &     &        &        & &    &   $H_{\alpha}$ morphology is asymmetric with hole in center\tablenotemark{31} \ \ ; \\
NGC4649........    & 15.92   &  -4.6   & S01(2)      & 6.50& 10.73  &   11.0 &\nodata&no\tablenotemark{9}    &  radio\tablenotemark{12} \ \ ;  $H_2$\tablenotemark{16} \ \ ;  HI\tablenotemark{17} \ \ ; no FIR dust\tablenotemark{25} \ \ ;\\
                   &         &         &             &     &        &        & &    &   $H_{\alpha}$ undetected at $log(L_{H_{\alpha}}) < 37.73$\tablenotemark{41} \ \ ; M60;   \\
NGC5018........    & 30.20   &  -4.5   & E3          & 3.81& 10.57  &   1.5  &\nodata&yes\tablenotemark{23}    &  radio\tablenotemark{12} \ \ ;  HI\tablenotemark{18} \ \ ; dust patches\tablenotemark{27} \ \ ;   \\
\\
NGC5044........    & 30.20   &  -4.8   & E0          &10.40& 10.70  & \nodata& yes\tablenotemark{1,11}&yes\tablenotemark{2}    &  nuclear radio source\tablenotemark{2} \ ; very bright, asymmetric  \\
                   &         &         &             &     &        &        & &    &  diffuse optical line emission; dusty region is\\
                   &         &         &             &     &        &        & &    &  smaller; bright dust cloud to East of center\tablenotemark{43} \ \ ;\\
NGC5831........    & 22.91   &  -4.8   & E3          & 3.11& 10.03  &    2.6 &\nodata&no\tablenotemark{5}    &      \\
NGC5866........    & 13.18   &  -1.3   & S03         & 4.98& 10.32  &    1.8 &\nodata&big disk\tablenotemark{23}    &  core radio source\tablenotemark{45} \ \ ; this S0 galaxy has a  \\
                   &         &         &             &     &        &        & &    &  strong dust lane along major axis\tablenotemark{27} \ \ ;\\
                   &         &         &             &     &        &        & &    &  contains HI; $H_{\alpha}$ emission\tablenotemark{47} \ \ ;\\
NGC6776........    & 70.41   &  -4.1   & E+          &9.21 & 10.66  &    3.2 &\nodata&\nodata&  a normal E1 with a very faint shell\tablenotemark{12, 13} \ \ \ \ \ ; \\
                   &         &         &             &     &        &        &       &       & 843MHz radio (very faint)\tablenotemark{26, 28} \ \ \ \ \ ; \\
NGC7562........    & 39.99   &  -4.8   & E2--3       & 4.65& 10.46  &   11.0 &\nodata&\nodata&  HI\tablenotemark{12, 19} \ \ \ \ \ ; a pair with NGC7562A and NGC5619 \\
                   &         &         &             &     &        &        & &    &  with possible tidal connection\tablenotemark{48} \ \ ; \\
                   &         &         &             &     &        &        & &    &  faint HI detected in old literature\tablenotemark{48} \ \ ;  \\
NGC7619........    & 39.99   &  -4.8   & E           & 6.20& 10.58  &    9.0 &\nodata&\nodata&  radio\tablenotemark{12} \ \ ; a normal E3\tablenotemark{27} \ \ ;   \\
NGC7626........    & 39.99   &  -4.8   & E           & 7.37& 10.61  &   11.7 &yes\tablenotemark{1}&yes\tablenotemark{8,9}    &  radio core-jet with double lobes \tablenotemark{8, 12} \ \ \ \ ; \\
                   &         &         &             &     &        &        & &    &  nuclear dust lane\tablenotemark{37} \ \ ;   \\
IC 1459 .......... & 18.88   &   -4.7  & E1          & 3.57& 10.37  & \nodata&yes\tablenotemark{1}&yes\tablenotemark{9}&  chaotic dust absorption visible; galaxy \\
                   &         &         &             &     &        &        & & &  crossed by a disk of ionized gas; \\
                   &         &         &             &     &        &        & & &  HI not detected, but CO ($H_2$) is\tablenotemark{49} \ \ ; heavy dust   \\
                   &         &         &             &     &        &        & & &  lane in $1^{\prime \prime}$ with more  chaotic and patchy dust \\
                   &         &         &             &     &        &        & &    &  further out; central starlight core is bluer\tablenotemark{50} \ \ ;\\
                   &         &         &             &     &        &        & &    &  ionized gas disk; stellar shells; bright point source \\
                   &         &         &             &     &        &        & &    &  in center; powerful compact radio core source\tablenotemark{51} \ \ ;\\

\enddata
\tablecomments{
Luminosities and distances are calculated with $H_0=75$ km s$^{-1}$ Mpc$^{-1}$;
$(^a )$  {data are from the LEDA catalog};
$(^b )$ {morphological type code from the LEDA catalog};
$(^c )$ {morphological type from the RSA catalog};
$(^d )$ {data are from \citet{ter02}};
$(^1 )$ {Goudfrooij et al. 1994a};
$(^2 )$ {Goudfrooij et al. 1994b};
$(^3 )$ {Raimond et al. 1981};
$(^4 )$ {Ebneter et al. 1988};
$(^5 )$ {Tran et al. 2001};
$(^6 )$ {Farrarese \& Ford 1999};
$(^7 )$ {Ferrarese, Ford \& Jaffe 1996};
$(^8 )$ {Verdoes Kleijn, et al. 2002};
$(^9 )$ {van Dokkum \& Franx 1995};
$(^{10} )$ {Zeilinger et al., 1996};
$(^{11} )$ {Ferrari et al. 2002};
$(^{12} )$ {Georgakakis et al., 2001};
$(^{13} )$ {Huchtmeier 1994};
$(^{14} )$ {Burstein, Krumm \& Salpeter 1987};
$(^{15} )$ {McNamara et al. 1994};
$(^{16} )$ {Sage \& Wrobel 1989};
$(^{17} )$ {Huchtmeier, Sage \& Henkel 1995};
$(^{18} )$ {Roberts et al. 1991};
$(^{19} )$ {Botinelli  et al. 1990};
$(^{20} )$ {Schweizer 1980};
$(^{21} )$ {Horellou et al. 2001};
$(^{22} )$ {Kim,  Fabbiano \& Mackie 1998};
$(^{23} )$ {Xilouris et al. 2003};
$(^{24} )$ {Leeuw, Sansom \& Robson 2000};
$(^{25} )$ {Temi et al. 2003};
$(^{26} )$ {NED};
$(^{27} )$ {Sandage \& Bedke 1994};
$(^{28} )$ {Mauch et al. 2003};
$(^{29} )$ {Schweizer 1980, 1981};
$(^{30} )$ {Goudfrooij et al. 2001; Whittmire et al. 2002};
$(^{31} )$ {Zeilinger et al. 1996};
$(^{32} )$ {Ferrari et al. 2002};
$(^{33} )$ {Mackie \& Fabbiano 1998};
$(^{34} )$ {Stevenson et al. 2002};
$(^{35} )$ {Giovvanini et al. 2001};
$(^{36} )$ {Ravindranath et al. 2001};
$(^{37} )$ {Jaffe et al. 1996};
$(^{38} )$ {Martel et al. 2002};
$(^{39} )$ {Chiaberge et al. 2003};
$(^{40} )$ {Marcum et al. 2001};
$(^{41} )$ {Ho, Filippenko \& Sargent 2003};
$(^{42} )$ {Biller et al. 2004};
$(^{43} )$ {Ferrari et al. 1999};
$(^{44} )$ {Filho et al. 2000};
$(^{45} )$ {Nagar et al. 2000};
$(^{46} )$ {Ho et al. 1997};
$(^{47} )$ {Plana et al. 1998};
$(^{48} )$ {Rembold et al. 2002};
$(^{49} )$ {Bettoni et al. 2001};
$(^{50} )$ {Carollo et al. 1997};
$(^{51} )$ {Forbes et al. 1995};
}
\end{deluxetable}





\begin{deluxetable}{lrrrc}
\tabletypesize{\tiny}
\tablewidth{0pt}
\tablecaption{Integrated Photometry}
\tablecolumns{9}
\tablehead{
\colhead{Name}                              &
\colhead{$F_{15\mu m}$\tablenotemark{a} }   &
\colhead{$F_{I}$\tablenotemark{a} }    &
\colhead{$F_{15\mu m}$/$F_{I}$}        &
\colhead{$\mu_{15\mu m}$ / $\mu_{I}$\tablenotemark{b}}     \\
\colhead{}                    &
\colhead{(mJy)}               &
\colhead{(mJy)}                    &
\colhead{}                    &
\colhead{}                     
}

\startdata

NGC0720........    & 6.8 $\pm$ 0.8  & 46.3 $\pm$ 11.3  & 0.15 $\pm$ 0.04  &   0.17     \\
NGC1316........    & 85.5 $\pm$ 7.4 & 157.8$\pm$ 22.9  & 0.54 $\pm$ 0.09  & \nodata    \\
NGC1399........    & 12.0 $\pm$ 1.3 & 78.0 $\pm$ 8.7   & 0.15 $\pm$ 0.03  &  0.16      \\
NGC1453........    & 4.2  $\pm$ 0.5 & 13.6 $\pm$ 3.6   & 0.31 $\pm$ 0.05  & \nodata    \\
NGC2300........    & 5.6  $\pm$ 0.8 & 36.5 $\pm$ 5.4   & 0.16 $\pm$ 0.04  &  0.16      \\
NGC4261........    & 15.3 $\pm$ 1.4 & 55.0 $\pm$ 7.1   & 0.28 $\pm$ 0.05  &  0.24      \\
NGC4278........    & 14.2 $\pm$ 1.2 & 55.1 $\pm$ 9.2   & 0.26 $\pm$ 0.05  &  0.25      \\
NGC4374........    & 27.6 $\pm$ 2.6 & 136.0$\pm$ 20.2  & 0.20 $\pm$ 0.04  &  0.19      \\
NGC4472........    & 59.6 $\pm$ 6.5 & 310.0$\pm$ 49.1  & 0.19 $\pm$ 0.03  &  0.16      \\
NGC4473........    & 7.0  $\pm$ 0.9 & 49.0 $\pm$ 7.1   & 0.14 $\pm$ 0.03  &  0.14      \\
NGC4552........    & 16.9 $\pm$ 2.1 & 74.0 $\pm$ 14.4  & 0.23 $\pm$ 0.05  &  0.17      \\
NGC4636........    & 37.0 $\pm$ 1.1 & 121.0$\pm$ 15.6  & 0.31 $\pm$ 0.04  &  0.29      \\
NGC4649........    & 35.8 $\pm$ 3.9 & 214.0$\pm$ 29.8  & 0.17 $\pm$ 0.03  &  0.17      \\
NGC5018........    & 11.8 $\pm$ 1.1 & 29.0 $\pm$ 6.0   & 0.41 $\pm$ 0.04  &  0.43      \\
NGC5044........    & 8.1  $\pm$ 0.9 & 42.0 $\pm$ 5.8   & 0.19 $\pm$ 0.04  &  0.20      \\
NGC5831........    & 3.5  $\pm$ 0.7 & 19.0 $\pm$ 3.3   & 0.18 $\pm$ 0.05  &  0.14      \\
NGC5866........    & 62.0 $\pm$ 3.2 & 63.0 $\pm$ 8.5   & 0.98 $\pm$ 0.14  &  1.10      \\
NGC6776........    & 5.7  $\pm$ 0.5 & 9.9  $\pm$ 2.2   & 0.58 $\pm$ 0.13  & \nodata    \\
NGC7562........    & 4.3  $\pm$ 0.5 & 19.8 $\pm$ 3.9   & 0.22 $\pm$ 0.05  & \nodata    \\
NGC7619........    & 5.7  $\pm$ 0.6 & 25.0 $\pm$ 4.1   & 0.23 $\pm$ 0.04  & \nodata    \\
NGC7626........    & 4.0  $\pm$ 0.7 & 20.4 $\pm$ 4.2   & 0.20 $\pm$ 0.05  & \nodata    \\
IC 1459........... & 12.0 $\pm$ 1.9 & 57.0 $\pm$ 6.2   & 0.21 $\pm$ 0.04  &  0.19      \\
\enddata

\tablecomments{
$(^a )$  {integrated flux inside $R_e$/8 aperture};
$(^b )$ {surface brightness ratio at $R_e$/4};
}

\end{deluxetable}

\end{document}